\documentclass[12pt,preprint]{aastex}

\shorttitle{Discovery of the first B[e] supergiants in M\,31}
\shortauthors{Kraus et al.}

\begin{document}

\title{DISCOVERY OF THE FIRST B[e] SUPERGIANTS IN M\,31$^*$}

\altaffiltext{*}{Based on observations
obtained at the Gemini Observatory, which is operated by the Association of Universities for
    Research in Astronomy, Inc., under a cooperative agreement with the
    NSF on behalf of the Gemini partnership: the National Science
    Foundation (United States), the Science and Technology Facilities
    Council (United Kingdom), the National Research Council (Canada),
    CONICYT (Chile), the Australian Research Council (Australia),
    Minist\'{e}rio da Ci\^{e}ncia, Tecnologia e Inova\c{c}\~{a}o (Brazil)
    and Ministerio de Ciencia, Tecnolog\'{i}a e Innovaci\'{o}n Productiva
    (Argentina), under program ID GN-2013B-Q-10.}

\author{M. Kraus}
\affil{Astronomick\'y \'ustav, Akademie v\v{e}d \v{C}esk\'e republiky,  
Fri\v{c}ova 298, 251\,65 Ond\v{r}ejov, Czech Republic}
\email{kraus@sunstel.asu.cas.cz}

\author{L. S. Cidale and M. L. Arias}
\affil{Departamento de Espectroscop\'ia Estelar, Facultad de Ciencias
Astron\'omicas y Geof\'isicas, Universidad Nacional de La Plata, B1900FWA, La Plata, Argentina \\ Instituto 
de Astrof\'isica de La Plata, CCT La Plata, CONICET-UNLP, Paseo del
Bosque s/n, B1900FWA, La Plata, Argentina}
 
\author{M. E. Oksala}
\affil{Astronomick\'y \'ustav, Akademie v\v{e}d \v{C}esk\'e republiky,
Fri\v{c}ova 298, 251\,65 Ond\v{r}ejov, Czech Republic}

\and

\author{M. Borges Fernandes}
\affil{Observat\'orio Nacional, Rua General Jos\'e Cristino 77,
20921-400 S\~ao Cristov\~ao, Rio de Janeiro, Brazil}

\begin{abstract}
B[e] supergiants (B[e]SGs) are transitional objects in the post-main 
sequence evolution of massive stars. The small number of B[e]SGs known 
so far in the Galaxy and the Magellanic Clouds indicates that this
evolutionary phase is short. Nevertheless, the strong aspherical 
mass loss occurring during this phase, which leads to the formation 
of rings or disk-like structures, and the similarity to possible 
progenitors of SN1987\,A emphasize the importance of B[e]SGs for the
dynamics of the interstellar medium as well as stellar and galactic 
chemical evolution. The number of objects and their 
mass loss behavior at different metallicities are essential ingredients for 
accurate predictions from stellar and galactic evolution calculations.
However, B[e]SGs are not easily identified, as they share many 
characteristics with luminous blue variables (LBVs) in their 
quiescent (hot) phase.
We present medium-resolution near-infrared $K$-band spectra for four stars in 
M\,31, which have been assigned a hot LBV (candidate) status. Applying 
diagnostics that were recently developed to distinguish B[e]SGs from hot LBVs, 
we classify two of the objects as bonafide LBVs; one of them currently in 
outburst. In addition, we firmly classify the two stars 
\object{2MASS\,J00441709+4119273} and \object{2MASS\,J00452257+4150346} as 
the first B[e]SGs in M\,31 based on strong CO band emission detected in
their spectra, and infrared colors typical for this class of stars.
\end{abstract}

\keywords{Circumstellar matter --- infrared: stars --- stars: early-type --- 
stars: massive --- supergiants}

\section{INTRODUCTION}

Massive stars ($> 8$\,M$_{\sun}$), though few in number, play a fundamental 
role in the evolution of their host galaxies. Via high-density stellar winds, 
these objects strongly enrich the interstellar medium with chemically processed 
material, and deposit large amounts of momentum and energy into their 
surroundings during their entire lifetime, until they explode as spectacular 
supernovae.
The post-main sequence evolution of massive stars contains several short-lived
transition phases, in which stars undergo strong massloss and mass ejections. 
The massloss during each phase greatly drives the subsequent evolution 
of the star 
\citep[e.g.,][]{1980A&A....90L..17M, 2005A&A...429..581M, 2012A&A...538L...8G}. 
However, the amount of mass lost in each transition phase is still highly 
uncertain as it depends on many parameters, such as stellar rotation, 
metallicity, magnetic fields, etc, several of which are still poorly 
constrained. Hence, for accurate predictions from stellar evolution 
calculations, detailed knowledge of the properties of these massive stars in 
transition is of vital importance \citep[e.g.,][]{1994A&A...290..819L, 
2008A&ARv..16..209P, 2012ARA&A..50..107L}.

Two types of massive stars in transitional evolutionary stages, which occupy 
and share the upper-left domain in the Hertzsprung-Russell (HR) diagram, are 
particularly puzzling: luminous blue variables (LBVs) and B[e] supergiants 
(B[e]SGs). Stars in both groups are rare 
\citep[e.g.,][]{2013arXiv1311.4792C}, indicating that these intermediate 
stages have rather short lifetimes. Nevertheless, these are extremely important 
objects as they undergo strong, often asymmetric massloss, driving the 
dynamics and chemical evolution of the interstellar medium, as well as dust 
formation in their circumstellar environments. The ejected 
material of LBVs usually accumulates in either elliptic or bipolar nebulae of 
ionized gas \citep[e.g.,][]{1995ApJ...448..788N, 2003A&A...408..205W}, often 
with indication for clumped wind structures \citep{2005A&A...439.1107D}, and 
cool dusty rings \citep{2003A&A...412..185C}. On the other hand, B[e]SGs have 
dense, high-ionized 
polar winds and equatorial rings or disks \citep{1986A&A...163..119Z}, which 
are detached from the star and consist of low-ionized or neutral atomic 
material \citep{2007A&A...463..627K, 2010A&A...517A..30K, 2012MNRAS.423..284A} 
and hot molecular gas and warm dust \citep[e.g.,][]{1988ApJ...334..639M, 
1996ApJ...470..597M, 2006ApJ...638L..29K, 2010AJ....139.1993K, 
2010A&A...512A..73M, 2012A&A...548A..72C, 2012A&A...543A..77W, 
2013A&A...549A..28K}. The mass ejection and disk formation mechanisms in both 
types of stars are still largely unknown, however, it has been 
suggested that binarity, rapid rotation, or stellar pulsations might 
play a significant role.

LBVs undergo episodic mass loss and eruptions. They typically occur in two 
`flavors', hot and cool \citep[following the notation 
of][]{2007AJ....134.2474M}. Cool LBVs are in `outburst', and as such undergo an 
excursion to the red side of the HRD, i.e., they appear to evolve from a 
hot into a cool supergiant and simultaneously brighten until they reach visual 
maximum when approaching the cool edge. At visual minimum, LBVs reside
on the blue side, where they are considered to be in their hot or `quiescent'
phase. These periodic outbursts (so-called S-Dor cycles) happen on timescales 
of months to years. On much longer timescales (of a thousand years), LBVs may 
experience phases of very strong or eruptive mass loss.
 
Several B[e]SGs are found to show some sort of variability and sudden, enhanced
mass loss similar to what is reported for LBVs 
\citep[e.g.,][]{2010A&A...517A..30K, 2012MNRAS.423..284A, 2012MNRAS.426L..56O, 
2012MNRAS.427L..80T, 2013arXiv1305.0459C}, suggesting an evolutionary link, 
with B[e]SGs as either the progenitors or the descendants of LBVs 
\citep[e.g.,][]{1996ApJ...468..842S}.  While evolutionary calculations 
fail so far to predict the B[e]SG phase, the enrichment of the disk material in 
$^{13}$CO, which reflects the stellar surface enrichment in $^{13}$C, suggests 
that most (if not all) B[e]SGs are in an evolutionary stage just beyond the 
main-sequence \citep{2013A&A...558A..17O}. No information on the $^{13}$C 
surface enrichment is available for LBVs, so that other criteria for a possible 
link need to be considered. The occurrence of the LBV phase in the evolution of 
massive stars depends on the stellar mass \cite[e.g.,][]{2011BSRSL..80..266M}. 
Stars with initial masses lower than $\sim 40$\,M$_{\odot}$ turn into LBVs as 
post-red supergiants, and it seems that these low-luminosity LBVs could 
directly explode as core-collapse supernovae of Type II 
\citep{2013A&A...550L...7G}. In this scenario, a direct LBV progenitor or 
descendant stage for B[e]SGs appears unlikely. However, a star in this mass 
range can pass through a B[e]SG phase before evolving into a red supergiant. 
Stars with higher initial masses turn into LBVs much earlier, i.e., already 
during their redward evolution. B[e]SGs lack extended large-scale structures 
that might be identified with LBV nebula remnants, rendering it 
questionable that they could be LBV descendants. On the other hand,
B[e]SGs have been suggested to be the predecessor of blue supergiants with 
bipolar ring nebulae, in which the hot molecular disk material has expanded 
and cooled and condensed into dust particles \citep{2007AJ....134..846S}. 
Two well-known representatives of such blue supergiants are SBW1 and 
Sher\,25, and both of them are in turn very similar to the precursor of SN1987A 
\citep[][respectively]{2013MNRAS.429.1324S, 2008MNRAS.388.1127H}. 
Notably, \citet{2013arXiv1305.0459C} recently suggested the B[e]SG S18 
in the Small Magellanic Cloud to be a viable SN1987A progenitor candidate. 
Hence, B[e]SGs might themselves be supernova progenitors without evolving 
through an LBV phase\footnote{We would like to stress that despite the fact 
that some B[e]SGs display variabilities that might be associated with LBV 
behavior \citep[such as S18, see, e.g.,][]{2013arXiv1305.0459C}, none of them 
was so far observed to perform S-Dor-like cycles, i.e., none appeared as a cool 
supergiant. Hence, they are (still) considered `real' (or typical) B[e]SGs.}.
Whether all high-luminosity B[e]SGs will directly evolve into
SN1987A progenitors, or whether maybe a subset might indeed become LBVs,
is currently unknown, but underlines the need for larger samples 
of these rare objects. 

Identification of members of these two groups can be quite difficult.
The best known sample of B[e]SGs to date is located in the Magellanic Clouds, 
while unambiguous classification of Galactic counterparts is usually hampered 
by the lack of proper distances (and hence luminosity values), so that they 
can be easily confused with other objects surrounded by dense disks, such as 
the Herbig Ae/Be stars, unless near-infrared spectra display emission from 
$^{13}$CO \citep[see][]{2009A&A...494..253K}. 
B[e]SGs are found exclusively on the luminous, blue side of the HR diagram, and 
their optical spectra are practically indistinguishable from those of hot LBVs, 
as both types of stars display strong line emission formed in their dense 
circumstellar matter \citep{1998ApJ...507..210K, 2007AJ....134.2474M, 
2012A&A...541A.146C}. Classification of bonafide LBVs without evidence of
a giant eruption is particularly difficult, and requires both long-term and 
multi-wavelength 
monitoring. Much effort has been undertaken in recent years to resolve 
massive star populations and, in particular, to identify LBVs and B[e]SGs
in the Milky Way and Local Group Galaxies \citep[e.g.,][]{2007AJ....134.2474M, 
2010MNRAS.405.1047G, 2010AJ....139.2330W, 2012A&A...541A.146C} 
for obvious reasons: (1) samples of these rare objects are increased, 
which is vital to improve our understanding of stellar evolution of massive 
stars in general, and to study evolutionary connections between LBVs and 
B[e]SGs in particular, (2) extragalactic B[e]SGs are easily distinguished 
from, e.g., Herbig stars due to well constrained distances, and hence 
luminosities, and (3) the occurrence of these particular evolutionary phases 
can be studied as a function of metallicity. Potential LBVs are identified by 
searching for stars that are spectroscopically indistinguishable from known 
LBVs. These stars are called \lq LBV-candidates\rq. However, given the 
difficulties in disentangling spectroscopically between hot LBV 
candidates and B[e]SGs, every hot LBV candidate is at the same time 
also a B[e]SG candidate.

In a recent near-infrared survey, \citet{2013A&A...558A..17O} reveal that 
hot LBVs and B[e]SGs have clearly distinct infrared characteristics. The two 
groups populate discrete regions in the $J-H$ versus $H-K$ color-color diagram, due to 
the larger amount of hot dust in the environment of B[e]SGs. Furthermore, while 
both types of stars display practically identical emission features in their 
$K$-band spectra, the study found CO emission only in stars classified as 
B[e]SGs, likely a consequence of their massive, stable dusty disks, which allow 
molecules to form at significant rates for easily detectable emission. Although 
a few B[e]SGs were found to lack CO band emission, the presence of these 
molecular bands in combination with near infrared colors typical for B[e]SGs 
provide clear evidence for a B[e]SG rather than an LBV nature of the 
star\footnote{The star HR\,Car is unique, as it is the only LBV with 
occasional, highly variable CO band emission \citep{1997ASPC..120...20M}. 
However, its $JHK$ magnitudes position it clearly in the LBV corner of the 
color-color diagram.}. These two characteristics hence seem to provide an ideal 
tool to separate B[e]SGs from hot LBVs, and we apply it  
to identify B[e]SGs in Local Group Galaxies, which are hidden in hot LBV 
candidate samples. In this Letter we present our first results on four objects 
in M\,31.

\section{OBSERVATIONS}

We selected four stars in M\,31 with 2MASS object names J00404307+4108459,
J00433308+4112103, J00441709+4119273, and 
J00452257+4150346. Three were identified by 
\citet{2007AJ....134.2474M} as hot LBV candidates, while 
J00433308+4112103 (= \object{AF\,And}) is a bonafide LBV 
\citep{1953ApJ...118..353H}. $K$-band spectroscopic observations of the 
targets were obtained in August and October 2013 at GEMINI North with the 
GEMINI Near InfraRed Spectrograph (GNIRS).
Using the longslit mode with the 0.\arcsec 30 \ slit, the short camera, and the 
110 l/mm grating, we achieved a resolving power of $\sim 5900$ and a wavelength
coverage from 2.23 to 2.40 $\mu$m.
Several ABBA sequences (nodding along the slit) were taken for each target.
An early A-type star (close to the target in position and airmass) was
observed immediately before or after each sequence for telluric absorption 
correction. Flats and arcs were taken with each source. We used 
IRAF\footnote{IRAF is distributed by the National Optical 
Astronomy Observatory, which is operated by the Association of Universities for 
Research in Astronomy (AURA) under cooperative agreement with the National 
Science Foundation.} software package tasks to extract and calibrate the 
spectra. The data reduction steps included subtraction of the AB pairs, 
flatfielding, telluric-correction, and wavelength calibration. Finally, the 
spectra were corrected for heliocentric and systemic velocities, 
and normalized. Observation details are listed in Table\,\ref{tbl_obs}.

\section{RESULTS}

\begin{figure}
\epsscale{0.85}
\plotone{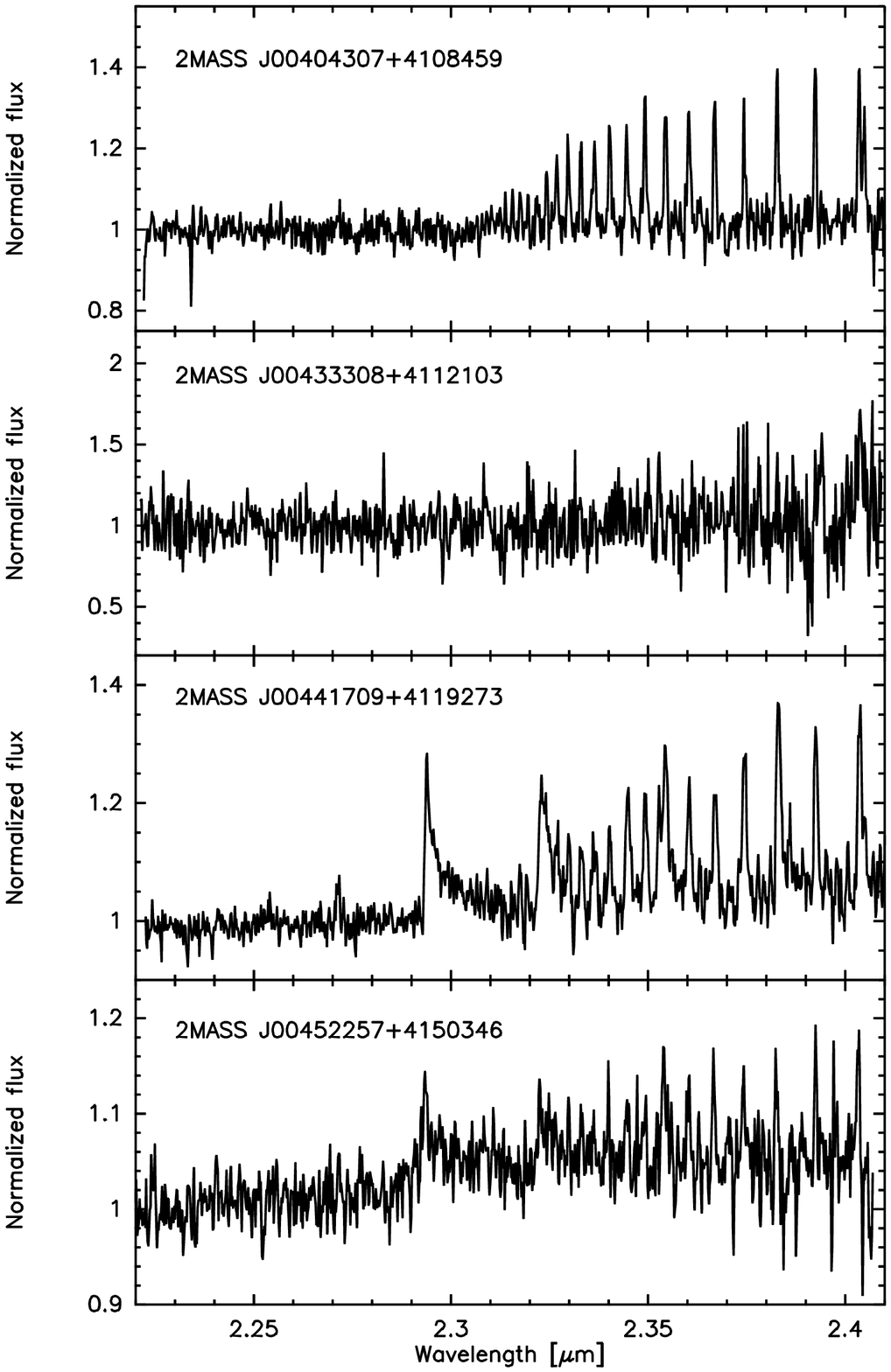}
\caption{Normalized $K$-band spectra of sample stars in M\,31 obtained
 with the GNIRS near-infrared spectrograph.}
\label{fig_spectra}
\end{figure}

\begin{figure}
\epsscale{0.85}
\plotone{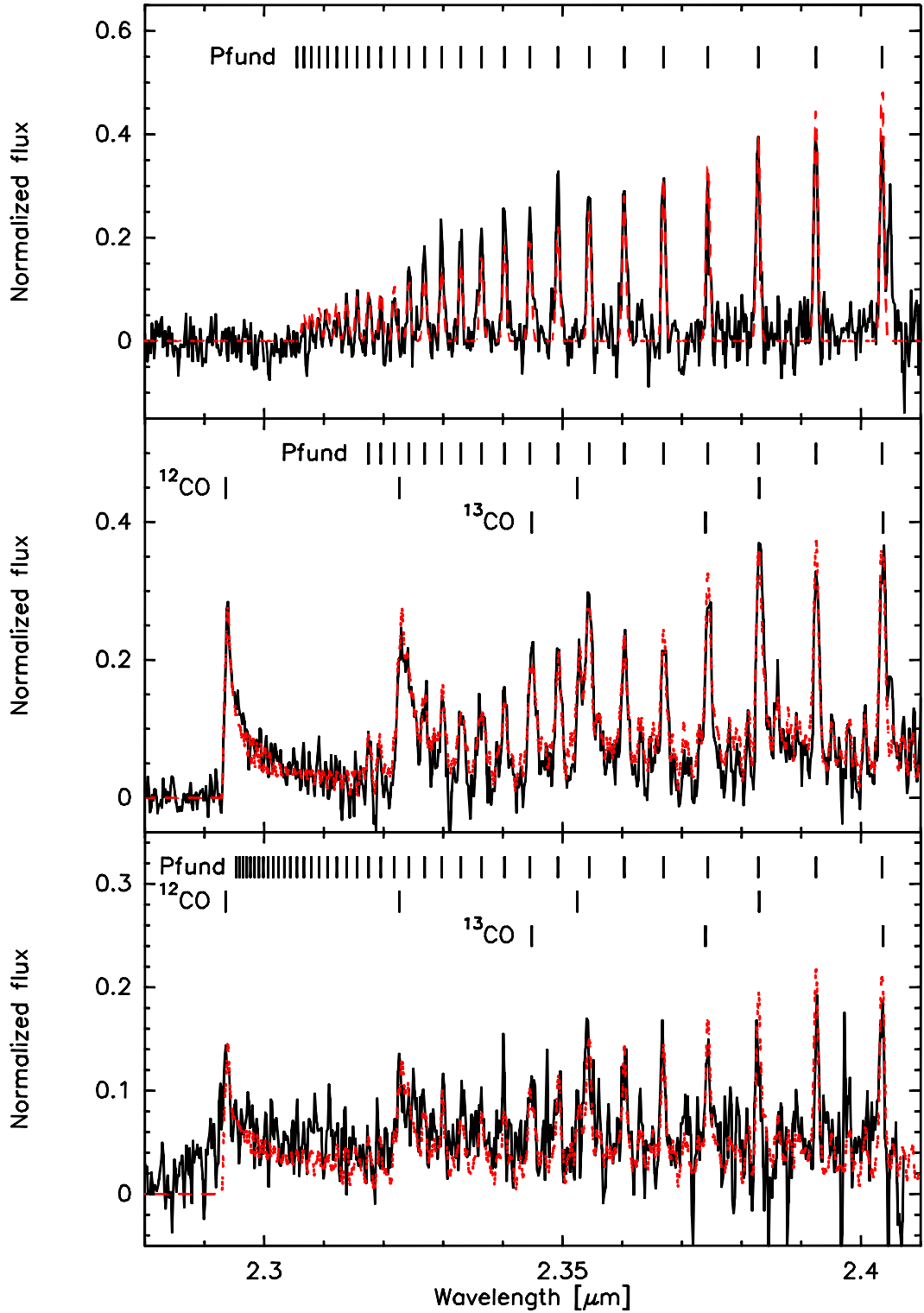}
\caption{Fit (grey/red in the online version) to the continuum subtracted 
Pfund line spectrum of
J00404307+4108459 (top) and to the CO band plus Pfund line spectra of
J00441709+4119273 (middle) and J00452257+4150346 (bottom). The ticks mark
the wavelengths of the hydrogen Pfund lines and of the $^{12}$CO and
$^{13}$CO band heads.}
\label{fig_co}
\end{figure}

The final normalized $K$-band spectra of the objects are shown in 
Fig.\,\ref{fig_spectra}. The spectra of 
J00441709+4119273 and J00452257+4150346 clearly display CO band 
emission and emission from the hydrogen Pfund series.
J00404307+4108459 only shows emission from the Pfund series. 
The spectrum of J00433308+4112103 is rather noisy, but 
there is no evidence for either CO band or Pfund line features.
We model the emission of the Pfund line series (Sect.\,\ref{pfund}) and the 
two molecular components $^{12}$CO and $^{13}$CO (Sect.\,\ref{co}); 
the best model fits to the observations are shown in Fig.\,\ref{fig_co}.

\subsection{Pfund Line Emission}\label{pfund}

Typically, emission from the hydrogen Pfund series originates from a 
dense wind. This high density environment can lead to pressure 
ionization effects, and hence to a sharp cut-off in the maximum number of
detectable Pfund lines. We model the Pfund recombination line spectra 
using the code developed by \citet{2000A&A...362..158K}. The line 
emission is not sensitive to the electron temperature. We therefore 
fix the temperature at $T_{\rm e} = 10\,000$\,K, 
which is a typical value for an ionized wind. The line profiles show no 
indication for rotational broadening beyond the spectral resolution of $\sim 
50$\,km\,s$^{-1}$. We thus adopt a pure Gaussian profile, which is a 
reasonably good approximation for optically thin recombination lines 
formed in a stellar wind. 

\subsection{CO Band Emission}\label{co}

We model the emission from the CO first-overtone bands
using the code developed by \citet{2000A&A...362..158K}, 
extended by \citet{2009A&A...494..253K} and \citet{2013A&A...558A..17O}. The
appearance of the band heads is caused by the superposition of many individual
ro-vibrational CO lines. Profiles of the ro-vibrational lines are typically
double-peaked, indicating that the emission originates from a (Keplerian)
rotating disk or ring, and the first CO band head, appearing at 2.3\,$\mu$m,
displays a characteristic structure consisting of a blue shoulder and a red peak
\citep[see, e.g.,][]{2000A&A...362..158K}. In the spectrum of
J00441709+4119273, the first CO band head lacks this characteristic structure
indicative of rotational broadening. Consequently,
any present rotational velocity, projected to the line of sight, must
be comparable or smaller than the spectral resolution, and we obtain a
reasonably good fit for $v_{\rm rot,CO} \sin i = 50\pm 5$\,km\,s$^{-1}$.
No satisfactory fit was obtained when using a pure Gaussian profile for the
individual ro-vibrational lines, excluding a spherical wind or shell as
the location of the CO molecules. The spectrum of J00452257+4150346 is too
noisy to make any inferences about the shape of its first band head, but, as the
spectral features look very similar to those in J00441709+4119273, we adopt the
same rotational broadening. From the strength of the higher band heads and 
the level of the quasi-continuum in between them, we obtain the molecular 
temperature and column density. The spectra also contain
emission from the molecular isotope $^{13}$CO,
and we determine a value of $7\pm 2$ for the $^{12}$CO/$^{13}$CO ratio
in both stars. This value mirrors the stellar surface enrichment in
$^{12}$C/$^{13}$C at the time of mass ejection. The complete set of model 
parameters derived from the fitting is given in Table\,\ref{tbl_param}. 
The CO emitting regions are confined to relatively cool but dense rotating
rings of material, detached from the surface of the star, as was also found
in Galactic and Magellanic Cloud B[e]SGs \citep[e.g.,][]{2010MNRAS.408L...6L,
2012A&A...548A..72C, 2013A&A...549A..28K, 2013A&A...558A..17O}.

\subsection{Color-Color Diagram}

\begin{figure}
\epsscale{0.85}
\plotone{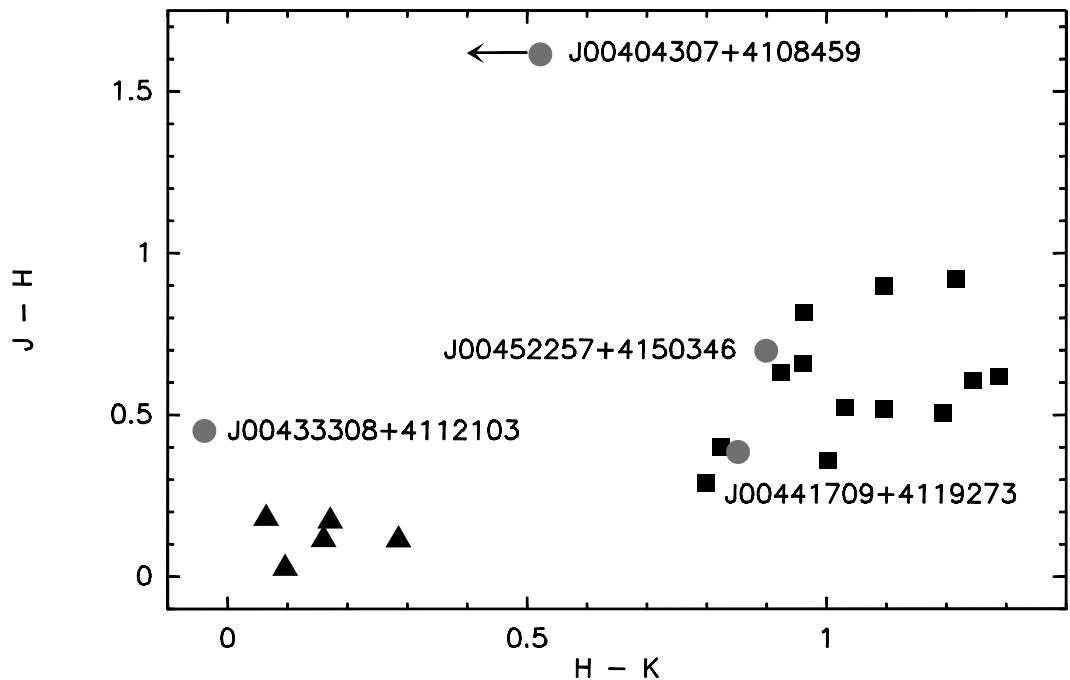}
\caption{J-H versus H-K color-color diagram. Values of known B[e]SGs
(squares) and LBVs (triangles) are taken from 
\citet{2013A&A...558A..17O}. Our sample stars are plotted as
circles. The H-K value of J00404307+4108459 is an upper limit;
its J-H value is uncertain.}
\label{fig_color}
\end{figure}

According to \citet{2013A&A...558A..17O}, LBVs and B[e]SGs are located in
distinct regions in the $J-H$ versus $H-K$ color-color diagram
(Fig.\,\ref{fig_color}). Using the colors of the sample stars, obtained from 
the $JHK$ magnitudes from the 2MASS point source catalog
\citep{2003yCat.2246....0C},
with the exception of J00452257+4150346 whose $J$ and $H$ band magnitudes are 
taken from \citet{2013ApJ...773...46H}, 
and plotting them together with the objects from \citet{2013A&A...558A..17O}
demonstrates that two objects, J00441709+4119273 and J00452257+4150346, 
reside in the region populated by B[e]SGs.
The star J00433308+4112103 falls into the LBV corner of the color-color 
diagram, confirming its bonafide LBV status.
The location of J00404307+4108459 is difficult to ascertain as
for this star only the $K$-band magnitude is accurately known. Its $J$- and
$H$-band values, and hence $H-K$, are only upper limits. 
The $H-K$ value is smaller than those of typical B[e]SGs, indicating
that J00404307+4108459 might belong to the group of hot LBVs.

\section{DISCUSSION AND CONCLUSIONS}

We present near-infrared spectra of three hot LBV candidates and one 
bonafide LBV star in M\,31. The chosen spectral range covers the wavelength 
region of the CO bands. Although not all confirmed B[e]SGs do show
CO band emission, their presence is an encouraging indication of a B[e]SG 
\citep{2013A&A...558A..17O}. 
A further, necessary criteria for classification as a B[e]SG
is provided by the star's position in the $J-H$ versus $H-K$ color-color diagram,
in which LBVs and B[e]SGs populate clearly distinct regions.

The bonafide LBV star in our sample, J00433308+4112103, shows no indication 
for CO band emission, and its $JHK$-band magnitudes place it into the LBV corner 
of the color-color diagram, confirming its LBV status. Furthermore, while
\citet{2007AJ....134.2474M} caught the star in its hot phase, its current 
featureless near-infrared spectrum resembles that of LHA\,120-S\,155, an LBV in 
the LMC found in outburst \citep{2013A&A...558A..17O}, indicating that 
J00433308+4112103 is currently also in outburst. 

Not much is known at present about the object J00404307+4108459. Based on its 
optical spectra, \citet{2007AJ....134.2474M} classify it as hot LBV candidate. 
Its near-infrared spectrum clearly displays emission from the hydrogen Pfund 
series, but no indication for CO band emission. As not all B[e]SGs show CO band 
emission, the absence of CO bands alone does not prohibit per se a
B[e]SG nature of the object. However, considering the location of the star in 
the infrared color-color diagram, which falls clearly outside the 
typical region for B[e]SGs, we can classify this star as a hot LBV.

CO emission has been detected in the objects J00441709+4119273 and 
J00452257+4150346, suggesting that these stars are in fact B[e]SGs. In an 
earlier study, \citet{1998ApJ...507..210K} mention that the star 
J00441709+4119273 (which they named k350) shares many characteristics with 
B[e]SGs, but as their optical spectrum does not show forbidden emission lines, 
these authors tentatively classify J00441709+4119273 as an LBV candidate. In 
contrast, the higher quality optical spectrum of \citet{2007AJ....134.2474M} 
shows these forbidden lines, typical criteria for B[e]SGs. 
\citet{2007AJ....134.2474M} also saw forbidden line emission in the spectrum 
of J00452257+4150346. This object was recently included in the study of 
\citet{2013ApJ...773...46H}, who suggest that the star is a warm hypergiant of 
spectral type A1\,Ia, similar to the yellow hypergiants. However, yellow 
hypergiants with CO band emission are found in an area close to the LBVs 
in the color-color diagram, and usually lack Pfund line emission 
\citep{2013A&A...558A..17O}. Based on the clear presence of CO band emission 
in our near-infrared spectra, and the location of the stars in the B[e]SG 
domain of the color-color diagram, we firmly classify J00441709+4119273 and 
J00452257+4150346 as the first B[e]SGs identified in M\,31.

The discovery of B[e]SGs in M\,31 opens new perspectives in studying 
members of this enigmatic group, which seem to be linked to progenitors
of supernovae of type SN1987\,A. With a metallicity about twice 
solar \citep[e.g.,][]{2012ApJ...758..133S}, M\,31 is an ideal laboratory 
for testing massive star evolutionary models at high metallicity. The 
$^{13}$CO enrichment of the circumstellar material found from our analysis 
is stronger than that typically found in Galactic and Magellanic Cloud B[e]SGs 
\citep[e.g.,][]{2013A&A...558A..17O}. As the evolution of massive 
stars changes severely with metallicity and rotation speed
\citep{2005A&A...429..581M}, this might indicate that either these stars 
were initially rapidly rotating, or they are 
very massive, with surface enrichment strongly enhanced due to 
the increased mass loss at higher metallicity.

To improve our understanding of stellar evolution of massive stars and their 
mass loss behavior at different metallicities, it is essential to continue to 
resolve massive star populations and identify B[e]SGs and LBVs in other
Local Group Galaxies. Furthermore, the methodology used here to distinguish 
between types of pre-supernova 
objects will be a crucial tool to study these stars in more obscured
areas, such as the Galactic center or other highly reddened regions, where
optical observations are not possible. This will increase 
both our knowledge and sample size leading to better statistics and more 
reliable conclusions about the nature and origin of B[e]SGs and LBVs.

\acknowledgments

We thank the anonymous referee for valuable comments on the manuscript.
This research made use of the NASA Astrophysics Data System (ADS). M.K. 
acknowledges financial support from GA\v{C}R under grant number P209/11/1198. 
M.E.O. acknowledges the postdoctoral program of the Czech Academy of Sciences. The 
Astronomical Institute Ond\v{r}ejov is supported by the project RVO:67985815.
L.C. and M.L.A. acknowledge financial support from the Agencia de Promoci\'{o}n
Cient\'{i}fica y Tecnol\'{o}gica (prestamo BID PICT 2011/0885), CONICET
(PIP 0300), and the Programa de Incentivos G11/109 of the Universidad Nacional 
de La Plata, Argentina. Financial support for International Cooperation of the 
Czech Republic (M\v{S}MT, 7AMB12AR021) and Argentina (Mincyt-Meys, ARC/11/10) 
is acknowledged.

\clearpage
                                                                                
\begin{deluxetable}{lccr}
\tablecaption{Observations\label{tbl_obs}}
\tablewidth{0pt}
\tablehead{
\colhead{2MASS object name} & \colhead{K$_{\rm s}$ \tablenotemark{a}} &
\colhead{T$_{\rm exp}$} & \colhead{S/N} }
\startdata
J00404307+4108459 & 15.300 & 1.22\,h & 30 \\
J00433308+4112103 & 15.406 & 1.12\,h &  9 \\
J00441709+4119273 & 14.728 & 1.26\,h & 50 \\
J00452257+4150346 & 15.415 & 1.80\,h & 35
\enddata
\tablenotetext{a}{Magnitudes are from the 2MASS point source catalog
\citep{2003yCat.2246....0C}.}
\end{deluxetable}

\begin{deluxetable}{lcccccccc}
\tabletypesize{\scriptsize}
\tablecaption{Best fit model parameters\label{tbl_param}}
\tablewidth{0pt}
\tablehead{
 & \multicolumn{4}{c}{Pfund lines} & \multicolumn{4}{c}{CO bands} \\
\colhead{Object} & \colhead{$v_{\rm gauss,Pf}$} &
\colhead{$T_{\rm e}$} &
\colhead{n$_{\rm max}$} & \colhead{$n_{\rm e}$} & \colhead{$v_{\rm rot,CO}
\sin i$} & \colhead{$T_{\rm CO}$} & \colhead{$N_{\rm CO}$} &
\colhead{$^{13}$C/$^{12}$C} \\
 & (km\,s$^{-1}$) & (K) &  & (cm$^{-3}$) & (km\,s$^{-1}$) & (K) &
($10^{21}$\,cm$^{-2}$) & }
\startdata
J00404307$+$4108459 & $50\pm 5$ & 10\,000 & 47 & $1.8\times 10^{13}$ &
    --- & ---  & ---  & ---  \\
J00441709$+$4119273 & $70\pm 10$ & 10\,000 & 39 & $5.7\times 10^{13}$ &
    $50\pm 5$ & $1850\pm 150$ & $1.5\pm 0.5$ & $7\pm 2$ \\
J00452257$+$4150346\tablenotemark{a} & $70\pm 10$ & 10\,000 & $\sim 60$ &
   $\sim 4.3\times 10^{12}$ & $50\pm 5$ & $1850\pm 150$ & $1.3\pm 0.7$ & $7\pm 2$
\enddata
\tablenotetext{a}{Due to the poor quality of the spectrum, the derived values
are only rough estimates.}
\end{deluxetable}

\end{document}